\documentstyle[12pt]{article}

\def\be{\begin{equation}}
\def\ee{\end{equation}}
\def\bea{\begin{eqnarray}}
\def\eea{\end{eqnarray}}

\def\beq{\begin{equation}}   \def\eeq{\end{equation}}
\def\bea{\begin{eqnarray}}   \def\eea{\end{eqnarray}}

\newcommand{\matel}[3]{\langle #1|#2|#3\rangle}

\begin{document}
\begin{flushright}
UND-HEP-01-BIG\hspace*{.2em}03\\
\end{flushright}


\vspace{.3cm}
\begin{center} \Large
{\bf
{CHARM PHYSICS -- LIKE BOTTICELLI IN THE SISTINE CHAPEL}}
\footnote{Invited talk given at KAON 2001, the 3th International
Conference on CP Violation, Pisa, Italy,
June 12 - 17, 2001}
\\
\end{center}
\vspace*{.3cm}
\begin{center} {\Large
I. I. Bigi }\\
\vspace{.4cm}
{\normalsize
{\it Physics Dept.,
Univ. of Notre Dame du
Lac, Notre Dame, IN 46556, U.S.A.} }
\\
\vspace{.3cm}
{\it e-mail address: bigi@undhep.hep.nd.edu }
\vspace*{0.4cm}

{\Large{\bf Abstract}}
\end{center}
After sketching heavy quark expansions as applied to
heavy flavour decays I emphasize the relevance of
nonperturbative dynamics at the charm scale for exclusive
$b\to c$ modes. I address the issue of
quark-hadron duality for charm and discuss both the experimental
and theoretical status of $D^0-\bar D^0$ oscillations. Finally
I argue that comprehensive CP studies of charm decays provide novel
portals to New Physics and suggest benchmark figures for
desirable sensitivities.


\tableofcontents
\section{Introduction}
\label{INTRO}

Since the title of my talk is admittedly far from illuminating, let me
explain its meaning.  There is an unimpeachable reason to visit the
Sistine Chapel, namely to see
Michelangelo's frescoes. For me they realize the power of beauty.
Most of you who have been in the Sistine chapel will have forgotten --
or maybe never have noticed in the first place -- that halfway up the
sidewalls there are wonderful frescoes by other famous masters, namely
Botticelli, Perugino and others. They are not quite on the
same unique level of Michelangelo's frescoes, yet had they been at almost
any other place in the world, people would undertake pilgrimages to
see just them!

Now I can state my analogy: I concede that the fascination of charm
decays might not match that of beauty (or of strange) decays anymore than
Botticelli can match the power of Michelangelo (or Rafaello)!
Of course, Botticelli is still Botticelli,
i.e. a first-rate artist, but what about charm? After all, the
{\em weak} phenomenology that the {\em Standard Model} (SM) predicts for
charm  is on the dull side. I will argue that
future charm studies can provide us with first rate lessons on
fundamental dynamics taking my cue from the two items italized in
the previous sentence:
\begin{itemize}
\item
With the weak dynamics expected to be well known, detailed charm
studies provide us with a test bed for theoretical  QCD technologies.
\item
Since features specific to the SM make the
weak phenomenology dull, charm transitions allow a novel access to the
flavour problem  with most -- though not all -- experimental conditions
being a  priori favourable!
\end{itemize}

Accordingly my talk will focus on two topics:
(A) Heavy
quark expansions and quark-hadron duality at the charm scale
with a discussion of $D^0 - \bar D^0$ oscillations and
(B) Charm's promise of revealing New Physics mainly through
studies of CP violation.

\section{QCD Technologies}
\label{QCD}

While we have no solution of full QCD, we do have theoretical
technologies inferred from QCD for special situations that allow us to
deal with nonperturbative dynamics in an internally consistent way. Those
are chiral perturbation theory for pion and kaon dynamics and heavy quark
expansions (HQE). The latter apply to various aspects of the dynamics of
beauty hadrons and possibly of charm hadrons as  well. However
since the charm quark mass exceeds ordinary hadronic scales
by  a moderate margin only, one can expect at best a
semi-quantitative description there.

Simulating QCD on the lattice represents a technology of wide reach.
In principle lattice QCD could work its way up to the charm scale from
below. However the considerable advances achieved recently on the
lattice with respect to heavy flavour physics were not based on such a
`brute-force' approach, but on a judicious use of $1/m_Q$
expansions \cite{HASHIMOTO}.

Quark models are still very useful -- if proper judgement is used.
Subtle, yet relevant field theoretic features of QCD
entering in the operator product expansion (OPE) like scale dependance
are often not realized in quark models, unlike in HQE.

\subsection{Heavy Quark Expansions}

In HQE one describes an observable $\gamma$ for a hadron $H_Q$
-- be it a total rate or a distribution --
through an expansion in inverse powers of the heavy
{\em quark} mass obtained through an
operator product expansion (OPE) \cite{HQT}
constructed at small Euclidean space-time intervals
$m_Q$:
\beq
\gamma(E) = \sum _i c_i(\alpha _S, E) (\Lambda _i/m_Q)^i \; ;
\eeq
$E$ denotes the relevant energy scale. Dispersion relations
that have to be exact as long as QCD does not generate unphysical
singularities in the complex plane connect the coefficients
of the OPE with moments of the observable distributions in
Minkowski space. This is expressed through sum rules
\cite{OPTICAL,VADE},
which can generically be expressed by
\beq
\int  dEw(E) \gamma (E)|_{hadrons} =
\int  dEw(E) \gamma (E)|_{quarks}
\eeq
stating that the integral of such observable $\gamma$ weighted
by some function $w(E)$ has to be equal when expressed in terms of
hadronic or quark degrees of freedom. This is referred to as
(global) quark-hadron duality or duality for short.

Such methods are applied to {\em in}clusive transitions -- lifetimes,
semileptonic branching ratios, lepton spectra etc. -- and
{\em ex}clusive observables like semileptonic form factors.

Quark models are still the best we have for treating nonleptonic
two-body modes of charm mesons. I understand there are
several reasons why the recently suggested methods for
$B\to M_1M_2$ \cite{BENEKE} are hard to justify for charm decays; one
is that nonleading corrections $\sim {\cal O}(1/m_Q)$ cannot be
treated (yet).
Nevertheless one  should try them there anyway!

\subsection{Applications: Lifetimes}

The HQE yields a more successful description of the pattern in
the weak lifetimes of charm hadrons than could a priori be expected,
in particular since those lifetimes differ by more than an order
of magnitude between $\tau (D^+)$ being the longest and
$\tau (\Omega _c)$ the shortest. Since no new data on lifetimes were
presented at this meeting, let me just make a few comments
here; my more detailed evaluation can be found in \cite{BCP3}:

$\bullet$
The HQE provides an after-the-fact rationale for most phenomenological
concepts like Pauli Interference, Weak Annihilation (WA) etc. as
${\cal O}(1/m_c^3)$ effects.

$\bullet$
It makes more definitive statements about the weight of those
concepts. For example, WA has to be a {\em non}leading effect in
{\em meson} decays, although it could still be quite significant.

$\bullet$
An important quantity is the ratio
$\tau (D_s)/\tau (D^0)$. Its first measurement by E 687, the precursor
of FOCUS, gave the first experimental confirmation that WA is indeed
{\em not} a {\em leading} mechanism for generating the $D^+$-$D^0$
lifetime  difference. It also provided clear evidence that the $D_s$
lifetime  exceeds that of $D^0$ by a moderate amount. A new round of
very high statistics experiments  has begun. The world average
from last summer reads \cite{GOLUTVIN}
\beq
\tau (D_s)/\tau (D^0) = 1.18 \pm 0.02
\eeq
rather than the previous world average of $1.125 \pm 0.042$.
A new SELEX number is a bit lower:
$\tau (D_s)/\tau (D^0) = 1.145 \pm 0.049$; new measurements
from FOCUS and the beauty factories will be added soon. Anticipating
the new world average to settle in around 1.2, it confirms
that WA is not the leading source of lifetimes differences
among charm mesons; at the same time it shows WA to be still
significant at the about 20 \% level, as expected
\cite{DS}.
The apparent fact that due to WA 10 - 20 \% of {\em all} $D_s$ decays
are interfered away
should  leave some clear footprints in certain classes of exclusive
channels. This could be studied, e.g., by comparing Dalitz plots of
Cabibbo  suppressed $D^0$ and $D^+$ modes with Cabibbo allowed $D_s$
channels.

$\bullet$
In contrast to quark model treatments the HQE
allow to understand the {\em absolute} $D^0$ and
$D^+$ semileptonic branching ratios as due to
$1/m_c^2$ effects.

$\bullet$
Predictions on baryon lifetimes involve quark model
estimates of various expectation values and thus are subject
to large theoretical uncertainties.

$\bullet$
We need $\sim$ 10 \% measurements of both $\Xi_c^+$ and
$\Xi_c^0$ lifetimes. They could easily reveal systematic
problems in the HQE predictions and have a significant impact
on our understanding of {\em beauty} baryon lifetimes.

$\bullet$
The {\em ratios} of semileptonic branching ratios for {\em baryons}
do {\em not} reflect their lifetime ratios!

\subsection{Theoretical uncertainties}

Since no clear evidence for New Physics has been found yet
in charm transitions (see the
discussion below), it would be tempting to declare victory and move
on to presumably greener pastures. I want to list three reasons
why charm physics still merits our dedicated attention:
\begin{itemize}
\item
It can still provide us with new insights into the inner
workings  of QCD.
\item
It allows us to calibrate the theoretical
tools we are using in extracting CKM parameters in $B$ decays:
measuring both decay constants $f_D$ and $f_{D_s}$ accurately
and comparing them with unquenched lattice results will enable
us to predict $f_B$ with more
confidence; likewise a precise extraction of the form factors
in $D \to l \nu K/K^*/\pi /\rho$ and their $q^2$ dependance
will be of direct as well as indirect help in extracting
$|V(ub)|$ from $B \to l \nu \rho/\pi$.
\item
The third motivation is not obvious:
the relevant scale for the nonperturbative dynamics in
{\em exclusive} $b\to c$ modes is given by the charm mass.
In particular
in $B \to l \nu D^*$ the most relevant preasymptotic effects are
given by the expansion in $1/m_c$ rather than $1/m_b$. Also the $D^*$
width has an impact on the accuracy with which the formfactor
for
$B\to D^*$ even at zero  recoil can be predicted. A comprehensive
analysis of charm decays  can shed light on such dynamics.
\end{itemize}
The central issue here is that of {\em theoretical uncertainties}.
They are fed from some obvious sources -- namely numerical uncertainties
in input parameters like $\alpha _S$ -- and not so straightforward
ones reflecting more systematic uncertainties. Limitations to duality
belong to the latter \cite{MISHA,VADE}.

Duality is a concept dating back to the early days of quark models.
It is, however, rarely appreciated that
over the last several years it has
become a fairly precise concept in heavy flavour decays rather than the
qualitative one it used to be; it also has to be viewed in the context
of the paradigm that QCD is the theory of strong interactions. The
corrollary of the latter is the statement that even hadronic
observables can be evaluated exactly on the quark-gluon level
{\em provided} all possible corrections to the quark-parton result
are properly accounted for. Duality violations are thus due to
corrections that could not be included due to a limitation in the
{\em algorithm} employed.

The OPE has intrinsic limitations: when constructed in Euclidean
space it has no sensitivity to terms of the type, say,
exp$\{ -m_Q/\Lambda \}$. Such innocuous contributions turn into
`oscillating' terms $\sim$ sin$(m_Q/\Lambda )$ upon continuation to
Minkowski space. Therefore the OPE will in general not yield
correct predictions {\em point for point} in $m_Q$ (or $E$ etc.):
some averaging or `smearing' in that variable will be required;
i.e., {\em local} duality will in general not hold. Furthermore
the expansion even for Euclidean quantities is only asymptotic
in $1/m_c$ and thus has irreducible errors even in principle.

In summary: charm studies can serve as a microscope for duality .
{\em At
best} we will encounter sizeable  uncertainties; {\em at worst} we might
be forced to conclude  that duality is not operative yet at the charm
scale.

\subsection{A case study: $D^0 - \bar D^0$ oscillations}

Oscillations are described by the normalized mass and width
differences:
$x_D \equiv \frac{\Delta M_D}{\Gamma _D}$,
$y_D \equiv \frac{\Delta \Gamma}{2\Gamma _D}$.
The experimental landscape is summarized by
\cite{CLEOBELLEFOCUS}:
\bea
x_D &\leq& 0.03   \\
 y_D &=&
\left\{
\begin{array}{l}
(0.8 \pm 2.9 \pm 1.0) \% \; \; {\rm E791} \\
(3.42 \pm 1.39 \pm 0.74) \% \; \; {\rm FOCUS} \\
(1.16^{+1.67}_{-1.65})\% \; \; {\rm BELLE} \\
(- 1.1 \pm 2.5 \pm 1.4) \% \; \; {\rm CLEO}
\label{DATAOSC}
\end{array}
\right.
\\
y_D^{\prime} &=& (-2.5 ^{+1.4}_{-1.6} \pm 0.3)\% \; \;
{\rm CLEO}
\; .
\eea
$y_D^{\prime}$ is extracted from fitting
a general
lifetime evolution to
$D^0(t) \to K^+\pi ^-$ and depends on the
strong rescattering phase $\delta$ between $D^0 \to K^-\pi^+$ and
$D^0 \to K^+\pi^-$:
$y_D^{\prime}
= -x_D{\rm sin}\delta + y_D {\rm cos}\delta $.
Obviously all measurements are still consistent with zero.
Yet to judge how significant that statement is, we have to examine
what the SM expectations are.

With $D^0 \to f \to \bar D^0$ transition amplitudes being proportional
to sin$\theta _C^2$ one has $x_D, y_D \leq 0.05$; furthermore
in the limit of $SU(3)_{Fl}$ symmetry those amplitudes have to vanish.
However a priori one cannot count on that being a very strong suppression
for the real world; thus
$ x_D, \; y_D \sim {\cal O}(0.01)$ represents a
conservative SM bound. On general grounds I find it unlikely
-- though mathematically possible -- that New Physics could
overcome the Cabibbo bound significantly. Comparing this general bound
on the oscillation variables to the data listed
in Eq.(\ref{DATAOSC}), I conclude the hunt for New Physics realistically
has only just begun!

One can give a more sophisticated SM estimate for
$x_D$, $y_D$. There exists an extensive literature on it; however
some relevant features were missed for a long time. Quark box diagrams
yield tiny contributions only:
$
x_D({\rm box}) \sim {\rm few} \; \times 10^{-5}
$.
Various schemes are then invoked  to describe selected hadronic
intermediate states to guestimate the impact of long distance
dynamics:
$
x_D(LD), \; y_D(LD) \sim 10^{-4} - 10^{-3}
$.
Recently a new analysis
\cite{DOSC} has been given based on an OPE providing a
systematic treatment in powers of $1/m_c$, the GIM factors $m_s$
and the CKM parameters. It finds that the GIM suppression
by a factor of $(m_s/m_c)^4$, which is behind the result stated on
$x_D({\rm box})$
is {\em untypically severe} \cite{GEORGI}. There are contributions with
gentle GIM factors  proportional to $m_s^2/\mu _{had}^2$ or even $m_s/\mu
_{had}$.  They are due to higher-dimensional operators and thus
accompanied  by higher powers of $1/m_c$. Since those are not greatly
suppressed,  contributions of formally higher order in $1/m_c$ can become
numerically  leading if they are of lower order in $m_s$. These
terms are actually due to condensate terms in the OPE, namely
$\matel{0}{\bar qq}{0}$ etc. On the {\em conceptual} side we have
achieved significant progress: it is again the OPE that allows
to incorporate nonperturbative dynamics from the start in a
self-consistent way. {\em Numerically} there is no decisive change,
although the numbers are somewhat larger with a better
appreciation of the uncertainties:
\beq
x_D(SM)|_{OPE}, \; y_D(SM)|_{OPE} \; \sim \; {\cal O}(10^{-3}) \; .
\eeq

Yet despite the similarities in numbers for $x_D$ and
$y_D$ the dynamics driving these two
$\Delta C=2$ observables
are quite different:
\begin{itemize}
\item
$\Delta m_D$ being generated by contributions from virtual states
is sensitive to New Physics which could raise it to the
percent level. At the same time it necessarily involves an
integral over energies thus making it rather robust against
violations of local duality.
\item
$\Delta \Gamma _D$ being driven by on-shell transitions
can hardly be sensitive to New Physics. At the
same time, however,  it is very vulnerable to violations of local
duality: a nearby  narrow resonance could easily wreck any GIM
cancellation and raise  the value of $\Delta \Gamma _D$ by an order
of magnitude!
\end{itemize}

If data revealed $y_D \ll x_D \sim 1\%$ we would have a strong case to
infer the intervention of New Physics. If on the other hand
$y_D \sim 1\%$ -- as hinted at by the FOCUS data -- then two scenarios
could arise:
if $x_D \leq {\rm few}\times 10^{-3}$ were found, one would infer
that the $1/m_c$ expansion within the SM yields a correct
semiquantitative result while blaming the "large" value for
$y_D$ on a sizeable and not totally surprising violation of
duality. If, however, $x_D \sim 0.01$ would emerge, we would face a
theoretical conundrum: an interpretation ascribing this to
New Physics would hardly be convincing since $x_D \sim y_D$.
To base a case for New Physics solely on the observation of
$D^0 - \bar D^0$ oscillations is thus of uncertain value, unless
$x_D$ is found to exceed $y_D$ significantly!

\section{CP violation in charm decays}

Most of us view the SM as incomplete, and our efforts are focussed
on uncovering New Physics. Charm decays have
a good potential to reveal interventions of New Physics that
might not be manifest in beauty decays \cite{RIO}. For charm quarks are
the  only up-type quark allowing a full range of indirect searches
for New Physics. While $D^0 -\bar D^0$ oscillations are slow,
$T^0 - \bar T^0$ oscillations
cannot occur at all, nor can CP violation there, since
top
quarks decay before they can hadronize \cite{DOK}.
Direct CP violation can emerge in exclusive modes that command
decent branching ratios for charm, but are really tiny for
top with little coherence left. Finally charm decays proceed in an
environment populated with  many resonances which induce final state
interactions (FSI) of  great vibrancy. While this feature complicates the
interpretations  of a signal (or lack thereof) in terms of microscopic
quantities, it  is optimal for getting an observable signal. In that
sense it  should be viewed as a virtue rather than a vice.

Charm hadrons provide several practical advantages:
their production rates are relatively large; they possess long
lifetimes and
$D^* \to D\pi$ decays provide as good a flavour tag as one can have.
Charm transitions should thus be viewed as a
{\em unique} portal for studying the
{\em flavour} sector.

The most promising probe in such an enterprise is a comprehensive
search for CP violation. The data are summarized in Table \ref{CPAS}
\cite{PEDRINI,CLEOCP}.
\begin{table}
\begin{tabular} {|l|l|l|}
\hline
channel & World Average '00 & CLEO '01 \\
\hline
\hline
$D^0 \to K^+ K^-$ & $(0.5 \pm 1.6) \%$ &$ (0.1 \pm 2.2 \pm 0.8) \% $\\
\hline
$D^0 \to \pi^+ \pi^-$ & $(2.2 \pm 2.6) \%$ & $(2.0 \pm 3.2 \pm 0.8)\% $\\
\hline
$D^{\pm} \to K^{\pm} K^-\pi^+ $ & $(0.2 \pm 1.1) \%$ & \\
\hline
$D^0 \to K_S \pi^0$ & & $(0.1 \pm 1.3) \%$ \\
\hline
$D^0 \to \pi^0 \pi^0$ & & $(0.1 \pm 4.8)\% $\\
\hline
\end{tabular}
\centering
\caption{Data on direct CP asymmetries in $D$ decays}
\label{CPAS}
\end{table}
All numbers are still consistent with zero --
on the level of a few percent. This represents an impressive
increase in experimental sensitivity. Yet at the same
time I consider it unlikely (though not inconceivable)
that New Physics could induce CP asymmetries of 10 percent
or more. Therefore the search for CP violation
in charm transitions {\em only now} has entered a phase with real
promise.

\subsection{CP Violation -- Expectations}

\noindent {\em (i) Direct CP Violation in Partial Widths}

For an asymmetry to become observable
between CP conjugate partial widths, one needs two coherent
amplitudes with a relative {\em weak} phase and a nontrivial
strong phase shift.

In Cabibbo favoured as well as in doubly Cabibbo suppressed
channels those requirements can be met with New Physics only. There is
one exception to this general statement  \cite{YAMA}: the transition
$D^{\pm} \to K_S \pi ^{\pm}$ reflects the interference between
$D^{+} \to \bar K^0 \pi ^+$ and $D^+ \to K^0 \pi ^+$ which
are Cabibbo favoured and doubly Cabibbo suppressed, respectively.
Furthermore in all likelihood those two amplitudes will exhibit
different phase shifts since they differ in their isospin
content.
The known CP impurity in the $K_S$ state induces a
difference {\em without any theory uncertainty}:
\beq
\frac{\Gamma (D^+ \to K_S \pi ^+) - \Gamma (D^- \to K_S \pi ^-)}
{\Gamma (D^+ \to K_S \pi ^+) + \Gamma (D^- \to K_S \pi ^-)} =
-2{\rm Re}\epsilon _K
\simeq - 3.3 \cdot 10^{-3}
\label{DKSSM}
\eeq
In that case the same asymmetry both in magnitude as well
as sign arises for the experimentally much more challenging
final states  $K_L\pi ^{\pm}$.
If on the other hand New Physics is present in $\Delta C=1$ dynamics --
most likely in the doubly Cabibbo suppressed transition -- then both the
sign and the
size of an asymmetry can be different from the number in Eq.(\ref{DKSSM}),
and by itself it
would make a contribution of the {\em opposite} sign to the asymmetry in
$D^+ \to K_L\pi ^+$ vs. $D^- \to K_L\pi ^-$. An explicit model
by D'Ambrosio and Gao \cite{GAO} shows that a CP asymmetry
$\sim {\cal O}(1 \% ) $ could indeed be induced by New Physics through
the doubly Cabibbo suppressed amplitude that would have escaped
detection so far; it would also
affect $\Delta m_D$ only insignificantly!

Searching for {\em direct} CP violation in
Cabibbo suppressed $D$ decays as a sign for New Physics would
represent a very complex challenge: within the KM description one expects
to find asymmetries of order 0.1 \% \cite{TAUCHARM,NAPOLI};
yet it would be hard
to conclusively rule out some more or less accidental enhancement due to a
resonance etc. raising an asymmetry to the 1\% level.
Observing a CP
asymmetry in charm decays would certainly be a
{\em first rate discovery
irrespective of its theoretical interpretation}.
Yet to make a case that a
signal in a singly  Cabibbo suppressed mode reveals New Physics is
iffy. One has to analyze at least several channels
with comparable  sensitivity to acquire a measure of confidence in one's
interpretation.

\noindent {\em (ii) Direct CP Violation in Final State Distributions}

For channels with two pseudoscalar mesons or a pseudoscalar and a vector
meson a CP asymmetry can manifest itself only in a difference between
conjugate partial widths. If, however, the final state
is more complex -- being made up by three pseudoscalar or two
vector mesons etc. -- then it contains more dynamical information than
expressed by its partial width, and CP violation can emerge also through
asymmetries in final state distributions. One general comment
still applies: since also such CP asymmetries require the
interference of two weak amplitudes, within the SM
they can occur in Cabibbo
suppressed modes only.

In the simplest such scenario one compares CP conjugate
{\em Dalitz plots}. It is quite
possible that different regions of a Dalitz plot exhibit CP
asymmetries of varying signs that largely cancel each other when
one integrates over the whole phase space. I.e., subdomains of the
Dalitz plot could contain considerably larger CP asymmetries
than the integrated partial width.
Once a Dalitz plot is fully understood with all its contributions, one
has a  powerful new probe. This is not an easy goal to
achieve, though, in particular when looking for effects that  presumably
are not large. It might be more promising  as a practical matter to start
out with a more euristic approach.  I.e., one can start a search for
CP asymmetries by just looking at conjugate Dalitz plots. One simple
strategy would be to focus on an area  with a resonance band and analyze
the density in stripes {\em across} the  resonance as to whether there is
a difference in CP conjugate plots.

For more complex final states containing
four pseudoscalar mesons etc. other probes have to be
employed.  Consider, e.g.,
$
D^0 \to K^+K^- \pi ^+ \pi ^- \; ,
$
where one can form a T-odd correlation with the momenta:
$
C_T \equiv \langle \vec p_{K^+}\cdot
(\vec p_{\pi^+}\times \vec p_{\pi^-})\rangle
$.
Under time reversal T one has
$
C_T \to - C_T
$
hence the name `T-odd'. Yet $C_T \neq 0$ does not necessarily
establish T violation. Since time reversal is implemented
by an {\em anti}unitary operator, $C_T \neq 0$ can be induced by
FSI \cite{BOOK}. While in contrast to the situation
with partial width differences FSI are not required to produce
an effect, they can act as an `imposter' here, i.e. induce a T-odd
correlation with T-invariant dynamics. This ambiguity can unequivoally
be resolved by measuring
$
\bar C_T \equiv \langle \vec p_{K^-}\cdot
(\vec p_{\pi^-}\times \vec p_{\pi^+})\rangle
$
in $\bar D^0 \to K^+K^- \pi ^+ \pi ^- $; finding
$
C_T \neq - \bar C_T
$
establishes CP violation without further ado.

Decays of {\em polarized} charm baryons provide us with a
similar class of observables; e.g., in
$\Lambda _c \Uparrow \; \to p \pi ^+\pi ^-$, one can analyse the
T-odd correlation $\langle \vec \sigma _{\Lambda _c}
\cdot (\vec p_{\pi ^+} \times \vec p_{\pi ^-})\rangle$ \cite{BENSON}.

\noindent {\em (iii) CP violation involving $D^0 - \bar D^0$
oscillations}

The interpretation is much clearer for a CP
asymmetry involving oscillations, where one compares
the time evolution of transitions like $D^0(t) \to K_S \phi$,
$K^+ K^-$, $\pi ^+ \pi ^-$ \cite{BSD}
and/or
$D^0(t) \to K^+ \pi ^-$ \cite{KPI} with their CP conjugate
channels. A  difference for a final state $f$ would depend
on the product
\beq
{\rm sin}(\Delta m_D t) \cdot {\rm Im} \frac{q}{p}
[T(\bar D\to f)/T(D\to \bar f)] \; .
\eeq
With  both factors being
$\sim
{\cal O}(10^{-3})$ in the SM
one predicts a practically zero
asymmetry $\leq 10^{-5}$. Yet
New Physics could generate
considerably larger values, namely
$x_D \sim {\cal O}(0.01)$,
Im$\frac{q}{p}
[T(\bar D\to f)/T(D\to \bar f)] \sim  {\cal O}(0.1)$
leading to an asymmetry of ${\cal O}(10^{-3})$.
One should note that the
oscillation dependant term is linear in the
small quantity $x_D$ (and in $t$) --
$
{\rm sin}\Delta m_D t
\simeq x_D t /\tau _D
$ --
in contrast to $r_D$ which is
quadratic:
$
r_D \equiv \frac{D^0 \to l^-X}{D^0 \to l^+X}
\simeq \frac{x_D^2 + y_D^2}{2}
$.
It would be very hard to see $r_D = 10^{-4}$ in CP insensitive
rates. It could well happen that $D^0 - \bar D^0$
oscillations are first discovered in such CP asymmetries!

\section{Summary and Outlook}

We have learnt many important lessons from charm studies. Yet even so,
they do not represent a closed chapter. On one hand charm physics
can teach us many more important lessons about QCD and its
nonperturbative dynamics beyond calibration work needed for a
better analysis of beauty decays. On the
other it provides a unique portal  to New Physics through up-type quark
dynamics. In this latter quest only now have we begun to enter promising
territory, namely gaining sensitivity for $x_D$ and $y_D$ values of
order percent and likewise for CP asymmetries.

Without a specific theory of New Physics one has to strike a balance
between the requirements of feasibility and the demands of making a
sufficiently large step beyond what is known when advocating
benchmark numbers for
the experimental sensitivity. In that spirit I suggest the
following numbers:
\begin{enumerate}
\item
Probe $D^0 - \bar D^0$ oscillations down to
$x_D$, $y_D$ $\sim {\cal O}(10^{-3}) \hat = r_D \leq
{\cal O}(10^{-5})$.
\item
Search for {\em time dependant} CP asymmetries in
$D^0(t) \to K^+K^-$, $\pi ^+\pi ^-$, $K_S\phi$ down to the
$10^{-4}$ level and in the doubly Cabibbo suppressed mode
$D^0(t) \to K^+ \pi ^-$ to the $10^{-3}$ level.
\item
Look for asymmetries in the partial widths for
$D^{\pm} \to K_{S[L]}\pi ^{\pm}$ down to $10^{-3}$ and likewise
in a {\em host} of singly Cabibbo suppressede modes.
\item
Analyze Dalitz plots and T-odd correlations etc. with a sensitivity
down to ${\cal O}(10^{-3})$.
\end{enumerate}
Huge amounts of new information on charm dynamics will become available
due to data already taken by FOCUS and SELEX and being taken at the $B$
factories; there
is activity to be hoped for at Compass, BTeV and
LHC-B. And finally there are the activities that
could be pursued at a tau-charm factory at Cornell. We can be sure
to learn many relevant lessons from such studies -- and there may be
surprises when we least expect it.

\section*{Acknowledgments}
Many thanks go to Flavio Costantino and his team for organizing an
inspiring    meeting in the classy setting that Pisa (with some
assistance  from San Gimignano and Lucca) can provide
like very  few places in the world: the cathedral complex reminds us
for what kind of yardstick for our work we should ultimately aim.
This work has been supported by the NSF under the grants
PHY-0087419.

\end{document}